\documentclass{optica-article}

\journal{opticajournal} % for journals or Optica Open

\articletype{Research Article}

\usepackage{lineno}
\usepackage[utf8]{inputenc}
\usepackage{graphicx}
\usepackage{float}
\graphicspath{ {images/} }
\begin{document}

\title{Strong coupling of dipole trapped atomic ensembles to a ring cavity}

\author{T. Olney, V. Naniyil, V. Boyer, J. Goldwin\footnote{Present address: Quantinuum, Golden Valley, MN, USA} and A. B. Deb\authormark{*}}

\address{School of Physics and Astronomy, University of Birmingham, Birmingham B15 2TT, United Kingdom
}

\email{\authormark{*}a.b.deb@bham.ac.uk} %% email address is required; see note below about the corresponding author designation

% use {asbstract*} to suppress the copyright line. Copyright information will be added in production

\begin{abstract*} 
Cavity quantum electrodynamics systems using atoms in resonant optical cavities are central elements of many applications such as quantum
networks and quantum-enhanced sensing. We present a novel experimental setup that achieves strong spatial mode-matching between a resonant mode of a triangular ring resonator and an ensemble of ultracold atoms trapped in an optical dipole trap. We realise a large-volume, 0.5\,mK deep cavity-assisted optical dipole trap from a laser beam
of modest power locked to a resonant mode of the cavity through the intracavity power build-up effect, allowing efficient loading of the trap from a magneto-optical trap. We observe dipole trapping of atoms through the vacuum Rabi splitting of the transmission spectrum of a weak probe beam near resonant with the cavity, which demonstrates collective strong coupling of the ensemble of atoms to the cavity after the magneto-optical trap is turned off. The work paves the way for a highly suitable platform for cavity-assisted quantum sensing of magnetic and electric fields. 

\end{abstract*}

%%%%%%%%%%%%%%%%%%%%%%%%%%  body  %%%%%%%%%%%%%%%%%%%%%%%%%%
\vspace{8mm}
An atom coupled to a resonant mode of a high-finesse cavity provides the backbone of many groundbreaking experiments laying a foundation for the science of cavity quantum electrodynamics (cQED)\cite{ Goy1983, Gleyzes2007, Thompson1992}. The strong light-atom interaction in this system enables highly efficient mapping of states of the atoms to the light field, and vice versa. The key parameter representing the strength of the interaction between a single atom and a radiation field, and thereby the light-atom state inter-mapping, is the cooperativity $\mathcal{C}$, which is the ratio of probabilities of an atom scattering into the driving cavity mode to all the other modes of the electromagnetic field. Compared to the free space, $\mathcal{C}$ is enhanced in a resonant optical cavity by the mean number of photon round trips inside the cavity. In the so-called strong coupling limit $\mathcal{C}\gg 1$, the coherent light-atom interaction dominates over dissipative processes leading to key resources for applications such as quantum gates  or quantum networks. 

Achieving very large $\mathcal{C}$ with a single atom in a cavity requires a very high cavity finesse and a very small mode volume, which poses many technical challenges \cite{Trupke07, Wilk2007}. An ensemble with a number of atoms $N\gg 1$ coupled to a resonant mode of a cavity provides a robust platform with less stringent requirements, where collective coherent exchange of photons between atoms in the ensemble and the radiation field leads to a strong enhancement of the cooperativity $\mathcal{C}_N = N\times \mathcal{C} $. The system has found numerous applications such as cavity cooling, quantum phase transitions, quantum simulations, quantum-enhanced metrology using spin-squeezing, quantum memory for light, cavity optomechanics and ultranarrow linewidth laser sources \cite{Tuchman06,Chen11,Ningyuan,a3117,a12066}. These experiments typically use atoms pre-cooled in a magneto-optical trap which is switched off during the experiment of interest. This limits the interrogation time  of the atoms due to their finite transit time through the cavity mode as they fall under gravity~\cite{Kruse03} . To overcome this, magneto-optically cooled atoms can be subsequently loaded into optical dipole traps or in atom chips \cite{Colombe2007}. Both single atoms and atomic ensembles have been trapped inside optical cavities, by capitalising on the cavity power build-up effect of an optical mode that has a frequency resonant with the cavity, but far off-resonant with respect to a relevant atomic transition \cite{Mosk01,Kruse03, Nagorny03}. In conventional standing wave cavities, atoms are only trapped in the antinodes of the optical modes. Due to the different wavelengths of the trapping and the probing modes, the spatial mode-matching between the two is not optimal \cite{Lee2014}. Cross dipole traps using intracavity power buildup in a running-wave cavity have been realised but atom-trapping zone only covered a small region of the available probe mode volume \cite{bernon11}. 

In this paper, we describe an experimental apparatus that is able to achieve collective strong coupling of potassium atoms optically trapped inside a triangular ring cavity. The experimental system features a scheme based on a ultra-stable transfer cavity, that we use to lock both the probe and the trap lasers. This ensure common mode noise rejection, and enables us to continuously and independently tune the probe and trap lasers. With this system, we demonstrate the trapping of millions of cold atoms in the trap, that has near-perfect geometrical overlap with the mode of the resonant probe. This results in collective strong coupling of atoms to the optical cavity, observed as a large vacuum Rabi splitting of $> 30$\,MHz, limited by the frequency scan range of the light source. The resulting collective cooperativity is $> 180$. A simple theoretical model confirms that this strong coupling is a direct consequence of the excellent spatial overlap between the light field and the atomic cloud. We demonstrate sufficiently long lifetime of the atoms in the cavity built-up dipole (CBD) trap, limited by technical noise arising from magnetic field switch off and backscattering of light inside the cavity.
\begin{center}
\begin{figure*}[t]
\includegraphics[width=1\textwidth]{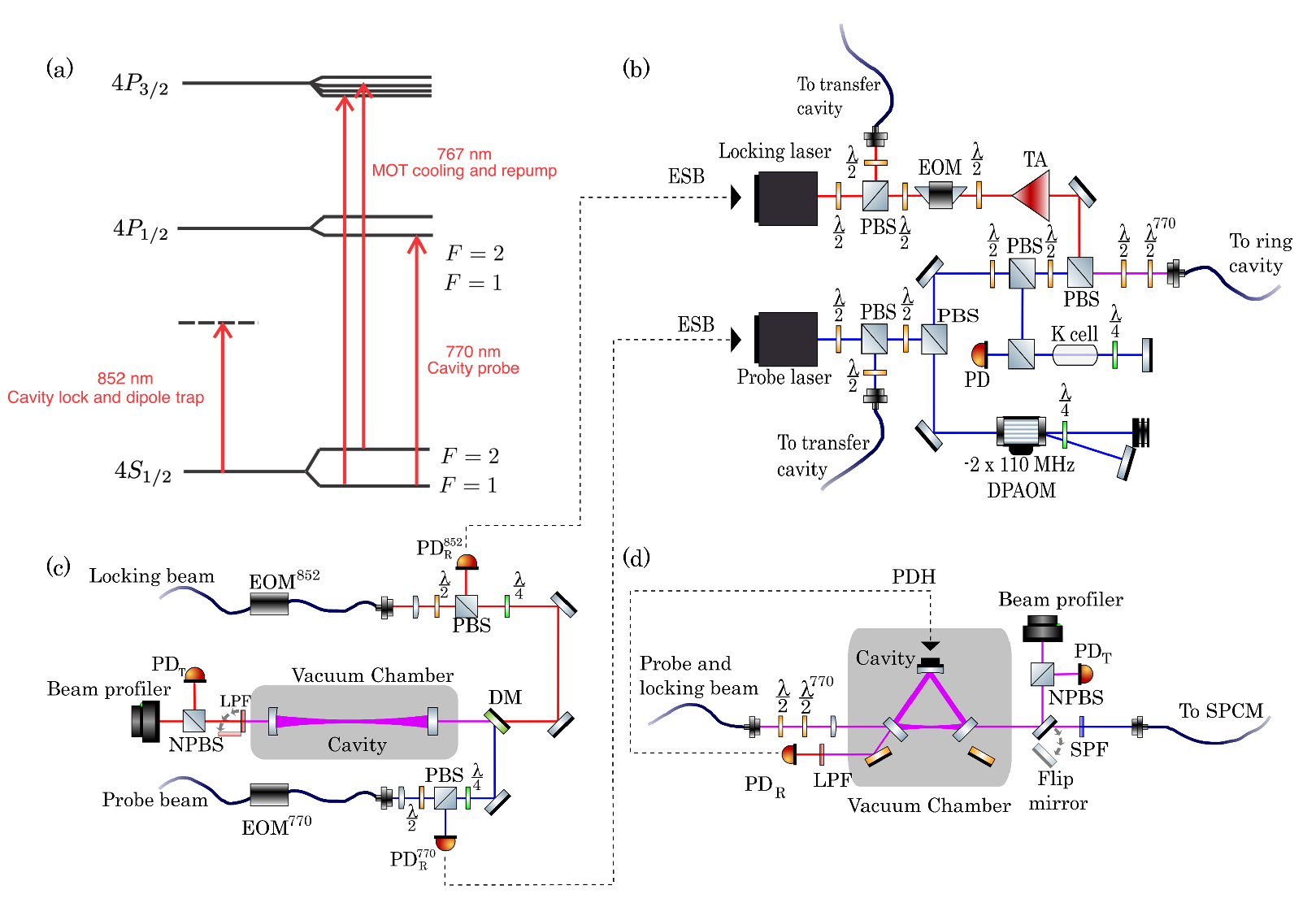}%
\caption{\label{fig:setup} a. The relevant atomic level structure and laser frequencies (not to scale). b. Optical layout of the cavity laser system. The locking laser operates at 852\,nm and the probe laser operates at 770\,nm. c. Optical layout for locking the cavity lasers. d. Optical layout for the ring cavity input and detection. Acronyms used are PBS: polarising beamsplitter, NPBS: non-polarising beamsplitter, $\frac{\lambda}{2}$: half waveplate, $\frac{\lambda}{4}$: quarter waveplate, EOM: electro-optic modulator, DPAOM: double-passed acousto-optic modulator, ESB: electronic sideband (lock), PDH: Pound-Drever-Hall (lock), TA: tapered amplifier, PD: photodiode, DM: dichroic mirror, LPF: long-pass filter, SPF: short-pass filter, SPCM: single photon counting module. Superscripts of $^{770, 852}$ indicate the component is narrowband at this wavelength, while subscripts of $_{T, R}$ indicate whether the photodiode is detecting transmissioin or reflection from the cavity. }%
\end{figure*}
\end{center}
\vspace{-8mm}

The starting point of our experiments is a magneto-optical trap (MOT) of $^{39}$K atoms produced using standard laser cooling techniques. We use cooling and repumping transitions near the D2 line of potassium at 767\,nm (Fig. 1a) and load $\sim 10^8$ atoms in a 3D MOT from a 2D MOT over 4\,~seconds. Following the initial loading, the atoms are further cooled and compressed leading to a final temperature of 200\,$\mu$K and a peak density of $3 \times 10^9$ cm$^{-3}$, with an $1/e^2$ diameter of 1\,mm.  

A triangular ring cavity, housed inside the 3D MOT chamber and comprising of two planar and one curved mirror, forms the basis of our experiments. The curved mirror is a plano-concave one with a radius of curvature of 100\,mm. It is glued to a piezoelectric element allowing the cavity length to be tuned. The total round trip cavity length is 95\,mm, resulting in a free spectral range (FSR) 3\,GHz. For s-polarised light, we measured the finesse $\mathcal{F}$ of the cavity to be 2100 at 770\,nm and 2400 at 852\,nm using the cavity ring down technique, resulting in a half-width at half-maximum linewidth (HWHM) of 800\,kHz and 700\, kHz, respectively. More details of our ring cavity are described in \cite{Culver16}. Good overlap of the MOT with the cavity is achieved by adjusting the magnetic fields and MOT beam alignment. The number of atoms in the MOT and their temperature are estimated from fluorescence imaging.

Fig.\,1(a) shows the relevant atomic level structure of potassium and the laser frequencies used in this work schematically. The laser used for locking the ring cavity operates at 852\,nm, far away from atomic resonances of potassium, whereas the probing laser operates near the D1 transition of potassium at 770\,nm.  A schematic of the cavity laser system is shown in Fig. \ref{fig:setup}(b). It comprises of two Toptica DLPro lasers, which are frequency-stabilised to an ultrastable reference cavity system (Stable Laser Systems Inc.)(Fig.\,1(c)), via electronic sideband locking (ESB) \cite{Thorpe08}, a modified Pound-Drever-Hall (PDH) technique \cite{Black01}. The reference cavity is 100 mm long with an FSR of 1.5 GHz. The cavity is made of one plano-concave mirror, with a radius of curvature of 50 mm, and one plane mirror. It is housed inside a vacuum chamber, and its mirrors and spacers are made of ultra-low expansion (ULE) glass. The whole vacuum system is temperature stabilised at the zero crossing temperature of the linear coefficient of thermal expansion of the ULE glass, which provides long term stability. The linewidth of the reference cavity was estimated from the transmission spectrum of laser beams through the cavity, taking into account the convolution of the laser and the cavity linewidths. For $852$ nm, the HWHM linewidth was estimated as $\approx 66$ kHz, while for $770$ this was estimated as $\approx 80$ kHz. These provide a lower bound to the cavity finesse of $11300$ and $9300$ for $852$ and $770$ nm light, respectively.

ESB locking provides a tunable lock which allows the science cavity and lasers to be stabilised at arbitrary, tunable detunings, and also results in low frequency drifts and common mode noise rejection \cite{Thorpe08}. In this scheme, a low frequency, PDH drive frequency $\nu_{\text{PDH}}$ is combined on a splitter with a tunable offset frequency $\nu_{\text{offset}}$. The combined radio frequency (RF) signal is used to drive an electro-optic modulator (EOM) creating frequency-sidebands on a carrier laser beam. The resultant laser light is comprised of a suppressed carrier beam with large sidebands at $\pm \nu_{\text{offset}}$, and each of these have small sidebands at $\pm \nu_{\text{PDH}}$. The frequency of the sideband at either $\pm \nu_{\text{offset}}$ is stabilised to the cavity mode, through standard PDH techniques, with the $\pm \nu_{\text{PDH}}$ sidebands providing the necessary phase information for the lock \cite{Black01}. The frequency at which the laser is stabilised can be tuned by varying $\nu_{\text{offset}}$. 

To create the PDH sidebands, around 1 mW of each of the beams is sent to a fibre-coupled EOM (Jenoptik models PM-830 and PM-770) before being coupled to the transfer cavity. The PDH frequency is provided by the Toptica PDD110f module, at 15.8 MHz for the 770 beam, and 17.3 MHz for the 852 beam. These signals are electronically amplified to a power level of a few dBm.  The offset frequency is provided by a rapidly configurable signal generator and amplified to 25 dBm provide larger offset sidebands. The offset frequency can be chosen at any frequency across the 1.5 GHz FSR of the reference cavity, meaning that the lasers can be tuned and stabilised at any frequency. 

While the locking scheme provides stable lasers in our scheme, the locking points are not related to an atomic reference. In this experiment, the probe light needs to be near resonance to the D1 $ F=1 \rightarrow F' = 2$ probing transition. Saturated absorption spectroscopy of potassiam D1 transition establishes this reference, prior to sending the light to the science chamber. The error signal for locking the laser is overlapped with the desired atomic transition by changing   $\nu_{\mathrm{offset}}$. We found that no adjustment to $\nu_{\mathrm{offset}}$  was necessary over a six month period of operation, demonstrating the long term stability of the reference cavity.

Figure \ref{fig:setup}(d) shows a simplified schematic of the input and detection optics for the locking and the probe lasers used for our ring cavity. Half wave plates and a plano-concave lens are used to match the polarisation and mode of the input light to the cavity. The reflected 852 nm light is used to lock the ring cavity via a standard Pound-Drever-Hall lock actuating the piezoelectric element. The transmitted light is spectrally filtered to remove the 852 nm light, and the remaining probe light is coupled to a single mode fibre. This light is then detected by a single photon counting module (SPCM), with a quantum efficiency of 65\% and a dark count rate of 250 counts per second (Perkin Elmer model SPCM-AQRH-13-FC).  In order to achieve fast scanning of the probe laser frequency, an additional acousto-optic modulator is used in a double-pass configuration. Scanning the probe laser frequency by tuning $\nu_{\text{offset}}$ should, in principle, be possible. However, imperfect mode-matching of the input beam to the cavity's TEM$_{00}$ mode introduced error signals from other modes in the transfer cavity. These spurious signals interfered with the TEM$_{00}$ error signal, ultimately causing the laser out of lock.

\begin{centering}    
\begin{figure}
\includegraphics[width=1\textwidth] {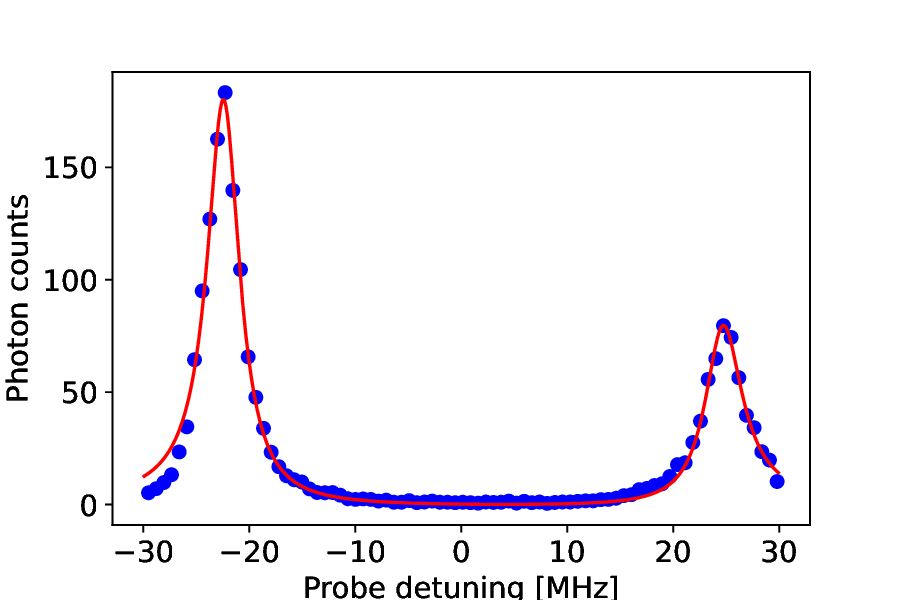}%
\caption{\label{avg_hist}  Vacuum rabi splitting for atoms in a MOT overlapped with the cavity mode. The MOT was loaded for 3 seconds. Each transmission spectrum was taken over 500\,$\mu$s immediately after the MOT lights were switched off, while the quadrupole magnetic field remained on. The blue points are the averages from 15 individual transmission spectra. The red line represents a fit to equation \ref{eq:transmission}, which gives an intracavity atom number of $N = (2.34 \pm 0.01) \times 10^5$. The asymmetry of the spectrum arises from a slight detuning of the cavity, of the order of a few MHz. }%
\end{figure}
\end{centering}

\begin{figure}
\includegraphics [width=0.8\textwidth]{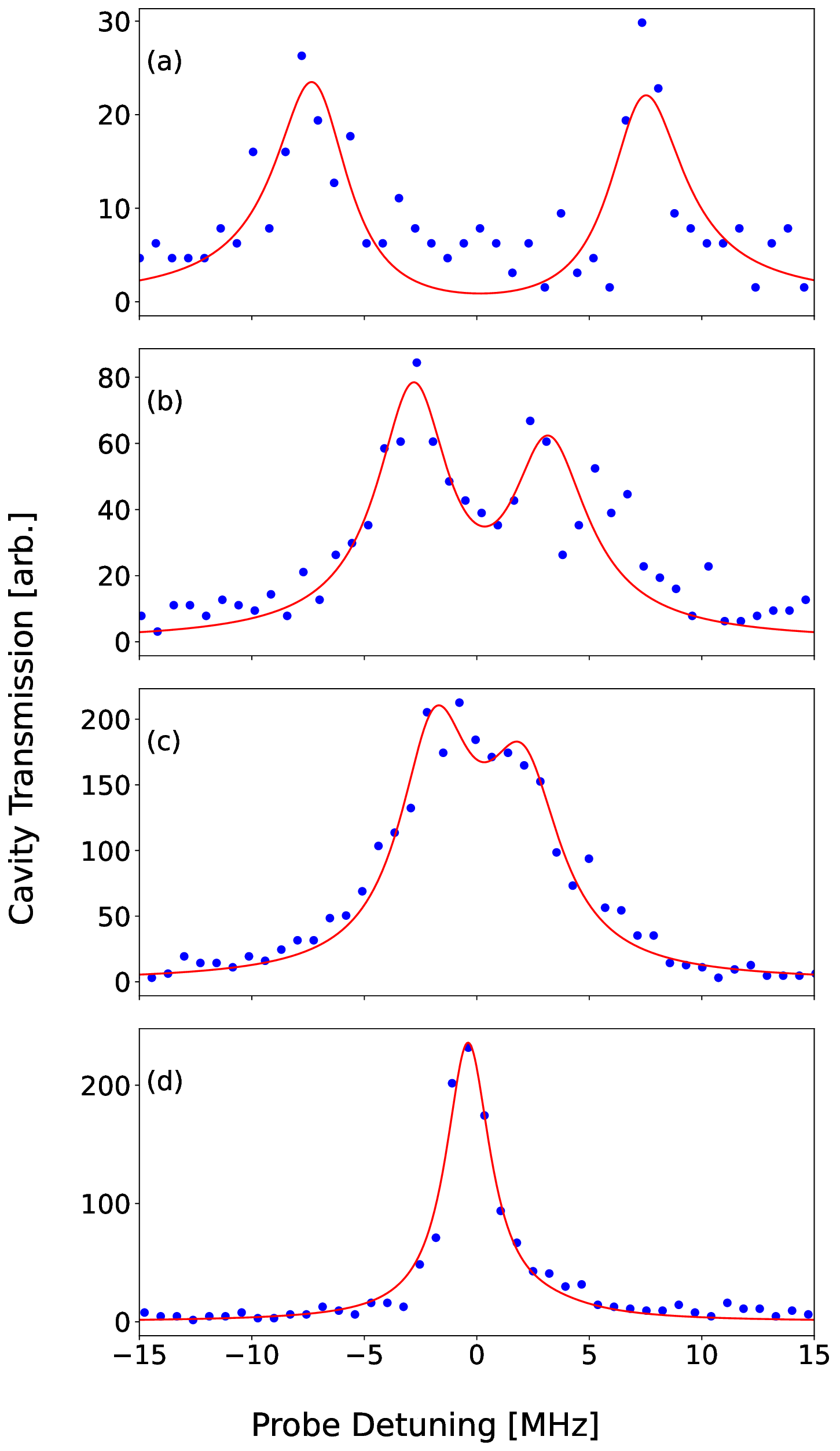}%
\caption{\label{dipole_trap_hold_times_2.eps} Vacuum rabi splitting for atoms in the cavity built-up dipole (CBD) trap at increasing hold times. The transmission spectra were taken over 500\,$\mu$s after the MOT lights and magnetic field were switched off. The spectras in (a)-(d) were observed 4\,ms, 16\,ms 30\,ms and 45\,ms after the magnetic field was ramped to zero. The blue dots represent the data from individual transmission spectra. The red lines represent a fit to equation \ref{eq:transmission}, which gives intracavity atom numbers of $N = (2.3 \pm 0.1) \times 10^4$, $(4.1 \pm 0.2) \times 10^3$, $(2.1 \pm 0.1) \times 10^3$, and $(7.4 \pm 4.4) \times 10^2$ for (a)-(d) respectively. }%
\end{figure}
For probing cold atoms in the cavity mode, we use a low power $\simeq$1\,nW probe light near resonant to the F=1 $\rightarrow$ F'=2 atomic transition coupled to the resonant TEM$_{00}$ mode of the cavity with an efficiency of 22\%. The resonant mode has a beam waist of \{90.2\,$\mu$m,  128\,$\mu$m\}. The probe is scanned across 60\,MHz centred on the atomic transition over 500 $\mu$s. The transmission spectrum was recorded as a histogram of photon counts on the SPCM. Figure 2a shows the probe beam transmission for atoms in a MOT after 3 seconds of loading.  A very clear normal mode splitting is evident in the spectrum, characteristic of strong collective coupling of atoms to the cavity. To quantify our observations, we fit our data to the cavity transmission described the cavity QED prediction \cite{Agarwal84}
\begin{equation} \label{eq:transmission}
    T = \frac{\kappa^2}{|\left( \kappa - i \Delta_c\right) +G^2/\left( \gamma - i \Delta_a \right)|^2},
\end{equation}
where $\kappa$ and $\gamma $ are the cavity and atomic decoherence rates respectively, $\Delta_c$ is the detuning between the probe laser and the empty cavity, $\Delta_a$ is the detuning between the probe and the atomic transition, and $G$ is the many-atom (collective) vacuum Rabi frequency. We use $G$ as a free parameter in our fit and subsequently the value for $G$  determines the number of atoms present in the cavity mode, $N$. The collective coupling $G$ can be expressed as $G=g\left(\xi N \right)^{1/2}$, where $\xi = 5/18$ is the relative oscillator strength averaged over all of the $F=1 \leftrightarrow F'=2$ transitions, and $g= 2\pi \times 91.5$\,kHz is the single atom vacuum Rabi frequency in this experiment. This results in an estimated $N = (2.34 \pm 0.01) \times 10^5$ atoms in the cavity mode. This compares favourably with the independently estimated number of atoms in the cavity mode $N = (1.0 \pm 0.2) \times 10^5$ obtained from atom number in the MOT measured from fluorescence images and the cavity mode volume. The discrepancy can be attributed to simplifying assumptions made during the estimation of the atom number from fluorescence imaging. For instance, the solid angle of light collected by the MOT camera was overestimated by assuming the full clear aperture of the imaging lens was utilised.

Central to this work is the collective coupling of atoms trapped in a CBD trap to the cavity.  To realise the CBD trap, we directed $200$\,mW of 852\,nm light to the ring cavity,  approximately 22\% of which is coupled to the TEM$_{00}$ mode of the cavity. This light is estimated to make roughly $\mathcal{F}/\pi \approx 600$ round-trips in the cavity, giving an intracavity power of $26$ W.  This provides a dipole potential minimum in which the potassium atoms can be trapped at the cavity waist, similar to a single free-running focussed Gaussian beam. With these parameters, we estimate a trap depth of 600\,$\mu$K for our CBD trap.  Figure\,3 shows several single-shot transmission spectra of near-resonant light through the cavity at various hold times in the CBD trap after the MOT is switched off.  This shows trapping of atoms in the CBD trap and their strong effective coupling to the cavity. In contrast, when the 852 nm beam's power was significantly reduced to an intracavity power of 0.5 W and a trap depth of 4 $\mu$K, we observed minimal to no atom trapping in the CBD trap, ruling out residual magnetic field gradients from the MOT coils as the cause of atom trapping. Due to constraints in putting high numerical aperture optics near the science chamber windows, we were unable to directly measure the number of atoms trapped in the CBD trap through fluorescence or absorption imaging. 

\begin{figure}
\includegraphics[width=\textwidth] {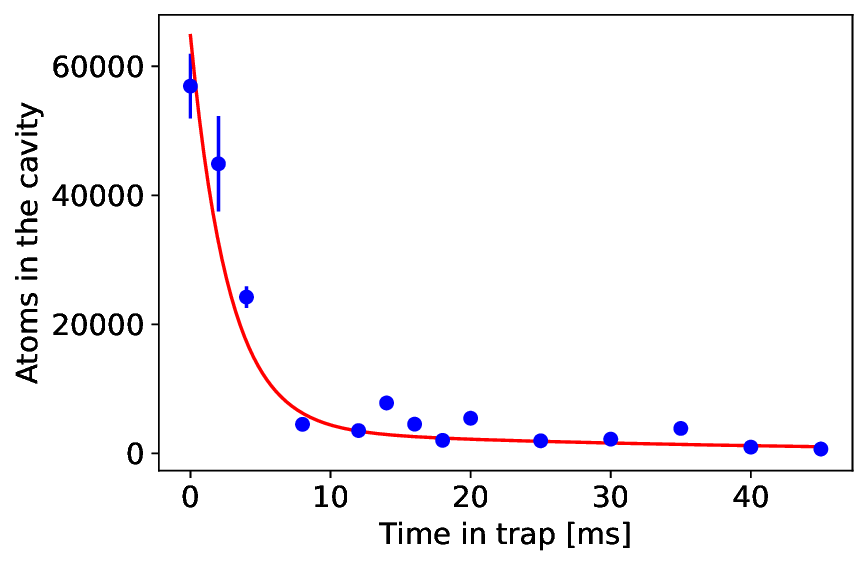}%
\caption{\label{lifetime} Lifetime of the cavity built-up dipole (CBD) trap. The blue dots are the data, which are the intracavity atom numbers averaged over 15 transmission spectra. The red line is a double exponential of the form $N(t) = N_0[\exp(-t/\tau_1) + \exp(-t/\tau_2)] $ fitted to the data. The fit gave decay times of $\tau_1 =3$ ms and $\tau_2 = 34$ ms for atoms in the CBD trap.  }%
\end{figure}

By looking at the amplitude of the normal mode splitting as a function of time, we can measure the lifetime of the atoms in the CBD trap. In Fig.\,~4, we plot the number of atoms in the CBD trap, extracted from the transmission spectra in Fig.~3, as a function of the hold time. In order to characterise the lifetime, we utilise a double exponential model of the form $N(t) = N_0[\exp(-t/\tau_1) + \exp(-t/\tau_2)] $ (the solid line in Fig.~4). This implies a fast loss of atoms initially with $\tau_1 = 3$\,ms, followed by a much slower loss rate of atom at later times with $\tau_2 = 34 $\,ms. These lifetimes are much shorter than usual far-off resonant dipole traps where the lifetime should be on the order of seconds given the trap parameters and the corresponding spontaneous photon scattering rate. This shortening of lifetime in CBD traps has been reported before in many previous experiments \cite{bernon11,Krenz07}, where it has been ascribed to the presence of backscattered fields within the cavity.  While the cavity mirrors are mostly reflective, some scattering of the forward propagating field is inevitable. Some of this scattered field can pump the degenerate back-propagating mode, leading to a power buildup and a standing wave of a modulation depth on the order of 1\%.  We observed evidence of a backward propagating field of this order by measuring the optical leakage power from one port of the ring cavity. Since the phase of the back-scattered field is random, the standing wave oscillates at a large range of frequencies, leading to rapid parametric heating of the atoms. It was shown in \cite{bernon11} that by minimising the effect of the backward propagating field, atom lifetime was increased from 100\,ms to several seconds. The relatively short lifetime in our trap, while non-ideal, is not an impediment for many future experiments in the setup, with the aim to perform cavity-enhanced magnetometry where a lifetime on the order of a millisecond is plenty as long as the atoms do not experience unwanted background magnetic fields, as is the case with our CBD trapped atoms. 

In conclusion, we have observed trapping of cold atoms in a CBD trap in a triangular macroscopic ring cavity. Strong collective coupling of these atoms to a resonant cavity mode was present in the form of large normal mode splitting of the transmission spectra of near-resonant light passing through the cavity. The lifetime of the atoms in the trap was found to be limited, owing to noise induced by the back-propagating intracavity scattered light field. By actively stabilising the phase of the backward propagating intracavity scattered light field, the lifetime could be increased to seconds. Our work paves the way for highly sensitive cavity-assisted quantum sensing, in particular cavity-enhanced magnetometry, where we aim to use a light beam to accrue a large dispersive phase shift from a large number of round-trips through a  CBD-trapped atomic ensemble acting as a birefringent medium. 

%%%%%%%%%% If using BibTeX:
%\bibliography{abcd}

\end{document}